\documentclass[13pt]{article} 
\usepackage{graphicx} 
\usepackage{color}
\usepackage{xcolor}
\usepackage{pdfsync}
\usepackage{amssymb}
\usepackage{amsmath,amsthm,amsfonts}  
\usepackage{graphicx}
\usepackage[breaklinks,colorlinks,backref]{hyperref}

\begin{document}

\title{
{\Large \bf Lorentzian Solitary Wave in a Generalised Nonlinear Schr\"odinger Equation}
}

\author{
M. A. Rego-Monteiro  \\
Centro Brasileiro de Pesquisas F\'\i sicas (CBPF)\\ 
Rua Xavier Sigaud 150, 22290-180 \\ Rio de Janeiro, RJ, Brazil \\
e-mail:   regomont@cbpf.br }  
\maketitle

\begin{abstract}
 
\indent

We obtain a travelling-wave solution of a generalised nonlinear Schr\"odinger equation with an additional term of the form $\Gamma(\psi(x,t)) =  \lambda \psi(x,t)^q$, where $\lambda$ and $q$ are real constants. Moreover, we show that  the density of energy of the system for the travelling-wave solution presents a Lorentzian solitary wave behaviour.

\end{abstract}

\begin{tabbing}

\=xxxxxxxxxxxxxxxxxx\= \kill

{\bf Keywords:} Nonlinear Schr\"odinger equation; solitary waves;  solitons; \\  Lorentzian solitary waves; Lorentzian solitons; \\ nonextensive thermostatistics.

 \\

\end{tabbing}


\section{Introduction}
\label{intro}
Nonlinear Schr\"odinger equations (NLSE) have been extensively used for the study of solitons in several areas of physics, as for instance nonlinear optics, plasma and water waves \cite{sulem, agrawal}. We show here that a Lorentzian is obtained as the energy density of a travelling solitary wave solution of the generalised NLSE \cite{q-3eq, qepl}
 (some authors call this generalised NLSE as NRT-nonlinear Schr\"odinger equation). It is interesting to notice that Lorentzian solitary waves were already found in structures with photonic band gaps \cite{contitrillo1,contitrillo2} and also in internal water waves \cite{benjamin, ono}.

From its proposal in \cite{q-3eq}, this generalised NLSE has attracted a lot of interest and since then some of its physical and mathematical aspects have been analised (\cite{pennini} - \cite{ ernesto}). The approach of classical field theory shows in \cite{qepl, jmp} that in order to describe the generalised NLSE it is necessary to consider an additional equation which is coupled to the previous one. Thus the generalised NLSE behaves as a master equation and there is an additional "slave" coupled equation that must be taken into consideration.

The approach developed in references  \cite{qepl, jmp} to solve the generalised NLSE was also shown to be useful in the study of a class of linear Hamiltonians. This class is chracterised by  non self-adjoint position-dependent mass Hamiltonians \cite{jpa}. These Hamiltonians had been previously discarded as being non-physical because they are not self-adjoint. However, in \cite{jpa} using the approach of classical field theory developed in \cite{qepl}, which leads to "master" and "slave" equations, was shown that it is always possible to find another self-adjoint Hamiltonian which is quantum mechanically equivalent to the non self-adjoint Hamiltonian considered in the class, thus restoring the physical status of these previously discarded Hamiltonians.

This generalised NLSE proposed in \cite{q-3eq} was inspired in a generalisation of Boltzmann-Gibbs statistical mechanics known as nonextensive statistical mechanics  \cite{tsallis1, tsallis2}. Several works have shown the consistency of this generalisation and its usefulness to describe several physical systems (see for instance \cite{tsallis2, tsallis3}).

In section \ref{rev}, we present a brief review of the generalised NLSE. In section \ref{lorentzian}, we obtain a solution of the generalized NLSE with the additional term $\Gamma(\psi(x,t)) =  \lambda \psi(x,t)^q$ and of its "slave" equation, where  $\psi(x,t)$ is the field present in that generalised NLSE, $\lambda$ and $q$ are real constants. We show that the density of energy presents a Lorentzian solitary wave behaviour for this solution. Moreover, we show the conservation of momentum. In section \ref{final} we present our final comments and we indicate the possibility of finding Lorentzian solitary waves in plasmas.

\section{The Generalised Nonlinear Schr\"odinger Equation}
\label{rev}

A key function in non-extensive statistical mechanics is a generalisation of the standard exponential function called $q$-exponential given as 
\begin{equation}
\label{qexp}
\exp_q (x) \equiv \left[ 1+(1-q) x  \right] ^{1/(1-q)} , \,\, q \, \epsilon \, \Re ,
\end{equation}
where $\exp_{q} (x) \rightarrow \exp (x)$, as $q \rightarrow 1$ , and $\exp(x)$ is the standard exponential function.

In \cite{q-3eq} a $d$-dimensional generalised NLSE was introduced for a particle of mass $m$,
\begin{equation}
\label{qnlse}
 i \hbar \partial_t  \left( \frac{\Psi(\vec{x},t)}{\psi_0} \right) = - \frac{1}{2-q} \frac{\hbar^2}{2 m} \nabla^2 \left( \frac{\Psi(\vec{x},t)}{\psi_0} \right)^{2-q} ,
\end{equation}
where $\partial_t \equiv \partial/\partial t$, $\nabla^2$ is the $d$-dimensional Laplacian and $\psi_0$ is a constant.
In the limit of $q \rightarrow 1$ equation (\ref{qnlse}) recovers the standard kinetic term of Schr\"odinger equation and it is easy to see that the generalisation of the plane wave
\begin{equation}
\label{qplanewave}
\Psi(\vec{x},t)= \psi_0 \exp_q(i (\vec{k} . \vec{x} - \omega t)) = \psi_0 \left[ 1+i(1-q) (\vec{k} . \vec{x} - \omega t)  \right] ^{\frac{1}{1-q}} , \,\, q \, \epsilon \, \Re ,
\end{equation}
is  a solution of the generalised NLSE given in eq. (\ref{qplanewave}) with
\begin{equation}
\label{qconditions}
\vec{p} = \hbar \vec{k} ; \, E = \hbar \omega ; \, E = \frac{\vec{p}\,^2}{2 m} \, .
\end{equation}
From now on, we use $\psi(\vec x,t) \equiv \Psi(\vec x,t)/ \psi_0$.

In \cite{qepl}  in order to have a consistent classical field theory for this model was shown that it is necessary to introduce a new field $\Phi(\vec x,t)$ which becomes the complex conjugate of $\psi(\vec x,t)$ only when $q=1$. Let us define the following Lagrangian
\begin{eqnarray}
\label{qLagrangian}
\mathcal L (\psi(\vec x,t),\Phi(\vec x,t)) = \left[ i \hbar \Phi(\vec x,t) \partial_t \psi(\vec x,t) - \frac{\hbar^2}{2m}  \psi(\vec x,t)^{1-q}  
\vec{\nabla} \Phi(\vec x,t) . \vec{\nabla} \psi(\vec x,t)  \right.  \nonumber \\
\left .    -  \Phi(\vec x,t) \Gamma(\psi(\vec x,t))  + c. \, c. \, t.              \right ]  ,
\end{eqnarray}
where $\Gamma(\psi(\vec x,t)) $ is a functional of $\psi(\vec x,t)$ and the $c. \, c. \, t.$  in the above equation indicates the complex conjugate terms of the three previous terms in the brackets. The Euler-Lagrange equations of the above Lagrangian for $\Phi(\vec x,t)$, $\psi(\vec x,t)$ and its complex conjugates  are
\begin{eqnarray}
\label{eulerlagrangepsi}
 i \hbar  \partial_t \psi(\vec x,t) &=& - \frac{1}{2-q} \frac{\hbar^2}{2m} \nabla^2  \psi(\vec x,t)^{2-q} + \Gamma(\psi(\vec x,t)) ,   \\
  \label{eulerlagrangephi} 
 -i \hbar  \partial_t \Phi(\vec x,t) &=& - \frac{\hbar^2}{2m} \psi(\vec x,t)^{1-q} \, \nabla^2  \Phi(\vec x,t) + \Phi(\vec x,t) \frac{\partial \Gamma(\psi(\vec x,t))}{\partial \psi(\vec x,t))}  \label{eulerlagrangephi} , \\
 -i \hbar  \partial_t \psi^{\star}(\vec x,t) &=& - \frac{1}{2-q} \frac{\hbar^2}{2m} \nabla^2  \psi^{\star}(\vec x,t)^{2-q} + \Gamma(\psi^{\star}(\vec x,t)) ,  \label{eulerlagrangepsistar} \\  
i \hbar  \partial_t \Phi^{\star}(\vec x,t) &=& - \frac{\hbar^2}{2m} \psi^{\star}(\vec x,t)^{1-q} \, \nabla^2  \Phi^{\star}(\vec x,t) + \Phi^{\star}(\vec x,t) \frac{\partial \Gamma(\psi^{\star}(\vec x,t))}{\partial \psi^{\star}(\vec x,t))}  . \label{eulerlagrangephistar}             
\end{eqnarray}
Note that, eq. (\ref{eulerlagrangepsi}) for $\Gamma(\psi(\vec{x},t))=0$ is the same as eq. (\ref{qnlse})  and there are additional equations for $\Phi(\vec{x},t)$ and its complex conjugate (eqs. (\ref{eulerlagrangephi}) and (\ref{eulerlagrangephistar})) which are linear equations  for $\Phi(\vec{x},t)$ when $\Gamma(\psi(\vec{x},t))=0$. The equation for $\Phi(\vec x,t)$ depends on the solution of eq. (\ref{eulerlagrangepsi}), i.e., it is a sort of "slave" equation. In order to find the solution of the system of equations we should  first obtain the solution of the equation for $ \psi(\vec x,t)$, (\ref{eulerlagrangepsi}), and insert this solution into equation for $\Phi(\vec x,t)$, in (\ref{eulerlagrangephi}). Moreover, it is clear by comparing eqs. (\ref{eulerlagrangephi}) and (\ref{eulerlagrangepsistar}) that, in general, $\Phi(\vec{x},t) \neq \psi^\star (\vec{x},t)$. Equation (\ref{eulerlagrangephi}) is linear when $\Gamma(\psi(\vec{x},t))=0$, while eq. (\ref{eulerlagrangepsistar}) is  nonlinear even for $\Gamma(\psi(x,t)=0$. 

This approach that considers a second field, namely $\psi(\vec x,t)$ and $\Phi(\vec x,t)$, has been shown interesting and useful even in the scope of linear systems. In \cite{jpa} a position-dependent mass quantum Hamiltonians (PDMH), that had been discarded because they were non self-adjoint, was  analysed. In order to construct a classical field theory for these systems was shown that it was necessary to introduce an additional field in the same way as it was introduced in the nonlinear system we presented above. However, in all PDMH considered in  \cite{jpa} there is always a transformation between the second field $\Phi(\vec x,t)$ and the complex conjugate of the first field, i.e., $\psi^\star(\vec x,t)$. This transformation is possible because for this linear system both equations for $\psi^\star(\vec x,t)$ and $\Phi(\vec x,t)$ are linear.  Moreover, with this transformation it is always possible to find another self-adjoint Hamiltonian, written in terms of $\psi(\vec x,t)$ and $\psi^\star(\vec x,t)$, which is quantum mechanically equivalent to the previous non self-adjoint Hamiltonian.

For the nonlinear system we are considering here it seems not to be possible to find a similar transformation to that described above for the linear system. Differently from the linear case, in the nonlinear case the differential equation for $\psi^\star(\vec x,t)$ is a nonlinear one for all values of $\Gamma(\psi(\vec{x},t))$, while $\Phi(\vec x,t)$ is described by a linear differential equation when $\Gamma(\psi(\vec{x},t))=0$. Thus, in case it exists, a transformation between $\Phi(\vec x,t)$ and $\psi^\star(\vec x,t)$ should transform a nonlinear differential equation into a linear one when $\Gamma(\psi(\vec{x},t))=0$.

\section{The Lorentzian Solitary Wave Solution}
\label{lorentzian}

From now on, we restrict our analysis to one spatial dimension. Let us now obtain a travelling-wave solution of the equations (\ref{eulerlagrangepsi}-\ref{eulerlagrangephi}) for $\Gamma(\psi(\vec x,t)) = \lambda \psi(\vec x,t)^q$. To this end it is convenient to define $z \equiv i (k x - \omega t)$. In terms of $z$ eqs. (\ref{eulerlagrangepsi}-\ref{eulerlagrangephi}) are written as
\begin{eqnarray}
 \hbar \omega  \psi^\prime(z) = \frac{\hbar^2 k^2}{2m} \big[ \psi(z)^{1-q}  \psi^{\prime \prime}(z) &+& (1-q) 
 \psi(z)^{-q} (\psi^{\prime}(z))^2 \big]+ \lambda  \psi(z)^{q} ,  \label{eulerlagrangepsiz} \\
 -\hbar \omega  \Phi^\prime(z) = \frac{\hbar^2 k^2}{2m} \psi(z)^{1-q}  \Phi^{\prime \prime}(z) &+& q \lambda \Phi(z)  \psi(z)^{1-q} \label{eulerlagrangephiz}.             
\end{eqnarray}
It is easy to verify that \cite{plastinosouza2014}) 
\begin{equation}
\label{solpsiz}
\psi(z) = \exp_q(z) 
\end{equation}
is a solution of (\ref{eulerlagrangepsiz}) where 
$ \lambda = \hbar \omega - \frac{\hbar^2 k^2}{2m}$. Let us consider the solution of eq. (\ref{eulerlagrangephiz}) for the above value of $\lambda$ and for the case where $\omega = \frac{c \hbar k^2}{2 m}$ with $c$ a real constant. In this case 
inserting also eq. (\ref{solpsiz}) into eq. (\ref{eulerlagrangephiz}) we find the following solution for $\Phi(z)$
\begin{equation}
\label{solphiz}
\Phi_{s}(z) = \exp_q(z)^{\frac{-c-q+a+1}{2}} \left[  c_1 + c_2 \exp_q(z)^{-a}  \right] \, ,
\end{equation}
where $c_{1,2}$ are two arbitrary constants and $a=\sqrt{-6q(1-c)+(1-c)^2+q^2}$.

It is easy to see that when $c \rightarrow 1$ eq. (\ref{solphiz}) becomes
\begin{equation}
\label{solphizpart}
\Phi_{c=1}(z) = c_1 \exp_q(z)^{-q} + c_2\, ,
\end{equation} 
which is the result found in \cite{qepl}.

The canonical conjugate fields to $\psi(x,t)$ and $\Phi(x,t)$ are
\begin{eqnarray}
\Pi_{\psi}(x,t) = \frac{\partial \mathcal{L}}{\partial \dot{\psi}(x,t)}  &=&  i \hbar \Phi(x,t)  \;  ;
 \Pi_{\psi^\star}(x,t) = \frac{\partial \mathcal{L}}{\partial \dot{\psi}(x,t)^\star} = -i \hbar \Phi(x,t)^\star   ;    \nonumber              \\
 \Pi_{\Phi}(x,t) = \frac{\partial \mathcal{L}}{\partial \dot{\Phi}(x,t)} &=& 0  \; ; \: \; 
 \Pi_{\Phi}(x,t)^{\star} = \frac{\partial \mathcal{L}}{\partial \dot{\Phi}(x,t)^{\star}} = 0  
 \label{momenta}.             
\end{eqnarray}
Using these above definitions with $\mathcal{L}$ given in eq. (\ref{qLagrangian}) we obtain for the Hamiltonian $\mathcal{H}$
\begin{eqnarray}
\label{densityhfinal}
\mathcal{H} = \frac{1}{2} \left[ \frac{\hbar^2}{2 m} \psi(x,t)^{1-q} \partial_x \Phi(x,t) \partial_x \psi(x,t)    + 
\lambda \Phi(x,t) \psi(x,t)^q + c.c.t. \right] \, .
\end{eqnarray}

Inserting the solutions $\psi_s(x,t)$ and $\Phi_s(x,t)$, the value of $\lambda= \hbar \omega-\frac{\hbar^2 k^2}{2 m}$  and choosing
\begin{equation}
\label{c}
\omega = \frac{c \hbar k^2}{2 m} 
\end{equation}
in $\mathcal{H}$ we get
\begin{eqnarray}
\label{densityhsolution}
\mathcal{E}(z) = \frac{1}{2} \frac{\hbar^2 k^2}{2 m} \left[  - \psi_s^{1-q}(z)  \Phi^{\prime}_s(z) \psi_{s} ^{\prime}(z)   + 
(c-1) \Phi_s(z) \psi_s(z)^q + c.c.t. \right] 
\end{eqnarray}
Consider for instance the solution $\Phi_s(z)$ given in eq. (\ref{solphiz}) with $c_1 = 0$ and $c_2 \neq 0$ \footnote{It may be easily verified that the same result is obtained by choosing $c_1 \neq 0$ and $c_2 = 0$ }. In this case we find
\begin{eqnarray}
\label{densityhsolution1}
\mathcal{E}(z) = \frac{c_2}{4} \frac{\hbar^2 k^2}{2 m} \left[ (3c+a+q-3) \exp_q(z)^{\frac{1+q-a-c}{2}}  \right] +  c.c.t. \; .
\end{eqnarray}
By choosing
\begin{eqnarray}
\label{lorentzianchoice1}
a &=& \frac{1+2q-q^2}{1+q}, \\
\label{lorentzianchoice2}
c &=& \frac{2}{1+q},
\end{eqnarray}
we obtain a Lorentzian solitary wave for the density of energy
\begin{eqnarray}
\label{lorentziandensityenergy}
\mathcal{E}(x,t) = \frac{2 c_2}{1+q} \frac{\hbar^2  k^2}{2 m} \frac{1}{1+(1-q)^2 (kx-\omega t)^2} \, .
\end{eqnarray}
Thus the energy of the solution in eqs. (\ref{solpsiz}) and (\ref{solphiz}) with constant $c$ in eq. (\ref{lorentzianchoice2}) is a Lorentzian solitary wave with total energy
\begin{eqnarray}
\label{totalenergy1}
E_s = \frac{2 c_2}{1+q} \frac{\hbar^2  k^2}{2 m} \frac{\pi}{| (1-q) k |} ,  \;\; q \neq 1 \;\; ,
\end{eqnarray}
where $| x |$ in the above equation means the absolute value of $x$. However, taking into account eqs. (\ref{c})  and (\ref{lorentzianchoice2}) and choosing $ c_2 = \frac{| (1-q) k |}{ \pi}$ the total energy of the Lorentzian wave is
\begin{eqnarray}
\label{totalenergy2}
E_s = \frac{2 }{1+q} \frac{\hbar^2  k^2}{2 m}  ,  \;\; q > -1, \; q \neq 1 \;\; .
\end{eqnarray}
It is important to stress that the density of energy is positive for all values of $x$ and the total energy is finite for this solution. Both are necessary conditions to have a sound solution.

Let us now show that the momentum density \cite{qepl}
\begin{eqnarray}
\label{densitycharge}
\rho_x(x,t) \equiv \frac{1}{2}  \left[ -i  \hbar \Phi(x,t) \partial_x \psi(x,t) + c.c.t.       \right]  ,
\end{eqnarray}
is a conserved charge of the system. Firstly, take the time derivative of $\rho_x(x,t)$
\begin{eqnarray}
\label{derivdensitycharge}
2 i \hbar \partial_t \rho_x(x,t) = i \hbar \partial_t \Phi(x,t) \partial_x \psi(x,t) + \Phi(x,t) \partial_x \left( i \hbar \partial_t \psi(x,t) \right)+ c.c.t.         .
\end{eqnarray}
Using the equations of this system, given in eqs. (\ref{eulerlagrangepsi} - \ref{eulerlagrangephi}), for $\Gamma(\psi) = \lambda \psi(x,t)^q$ one obtains
\begin{eqnarray}
\label{derivdensitycharge1}
2 i \hbar \partial_t \rho_x(x,t) = \frac{\hbar^2}{2 m}  \left[ \psi(x,t)^{1-q}  \partial_x \psi(x,t)  \partial_{x}^2 \Phi(x,t)  \right. \nonumber \\
- 3 (1-q) \Phi(x,t) \psi(x,t)^{-q}   \partial_x \psi(x,t) \partial_{x}^2 \psi(x,t) - \Phi(x,t) \psi(x,t)^{1-q} \partial_x^{3} \psi(x,t)  \nonumber \\
\left. + q (1-q) \Phi(x,t) \psi(x,t)^{-1-q}  \left( \partial_x \psi(x,t) \right)^3  + c. c. t.  \right]  .
\end{eqnarray}
Expanding the derivatives
\begin{eqnarray}
\label{derivativeterms}
\partial_x \left[   \Phi(x,t) \psi(x,t)^{1-q} \partial_x^2 \psi(x,t)    \right] ;   \;  \partial_x \left[  \psi(x,t)^{1-q}  \partial_x  \Phi(x,t)  \partial_x \psi(x,t)    \right] ; \nonumber \\
-(1-q) \partial_x \left[  \psi(x,t)^{-q}   \Phi(x,t) \left( \partial_x \psi(x,t) \right)^2   \right] ,
\end{eqnarray}
and identifying the expanded terms of these derivatives with the terms in eq. (\ref{derivdensitycharge1}) we obtain
\begin{eqnarray}
\label{conservation}
\partial_t \rho_x (x,t) + \partial_x j_x(x,t) =0 ,
\end{eqnarray}
where
\begin{eqnarray}
\label{current}
j_x(x,t) = \frac{\hbar}{4 m} \left[  - \Phi(x,t) \psi(x,t)^{1-q} \partial_x^2 \psi(x,t) +  \psi(x,t)^{1-q}  \partial_x  \Phi(x,t)  \partial_x \psi(x,t) \right. \nonumber \\
-  (1-q)  \psi(x,t)^{-q}   \Phi(x,t) \left( \partial_x \psi(x,t) \right)^2  - \Phi^{\star}(x,t) \psi^{\star} (x,t)^{1-q} \partial_x^2 \psi^{\star}(x,t)  \\
+ \left. \psi^{\star}(x,t)^{1-q}  \partial_x  \Phi^{\star}(x,t)  \partial_x \psi^{\star} (x,t) - (1-q)  \psi^{\star} (x,t)^{-q}   \Phi^{\star} (x,t) \left( \partial_x \psi^{\star} (x,t) \right)^2 \right] \nonumber
\end{eqnarray}
Finally, it is simple to realise that inserting the solutions eqs. (\ref{solpsiz}) and (\ref{solphiz}), for the values which give the Lorentzian solitary waves in eqs. (\ref{lorentzianchoice1}) and (\ref{lorentzianchoice2}), into the density eq. (\ref{densitycharge}) and integrating over all space we obtain
\begin{eqnarray}
\label{totalmomentum}
P_s = \hbar k \;\; .
\end{eqnarray}

\section{Final Comments}
\label{final}
We have found a travelling-wave solution of a generalised NLSE (\ref{eulerlagrangepsi}) with an additional term $\Gamma(\psi(x,t)) = \lambda \psi(x,t)^q$
and of its coupled equation (\ref{eulerlagrangephi}). The density of energy consistent with this generalised NLSE (\ref{eulerlagrangepsi}) is obtained from classical field theory (\ref{densityhfinal}).  For specific values of the constants of this solution, the density of energy (\ref{densityhsolution}) behaves as a Lorentzian solitary wave (\ref{lorentziandensityenergy}) of  total energy $E_s = \frac{2 }{1+q} \frac{\hbar^2  k^2}{2 m}$, with $q >-1, q \neq 1$, and momentum $P_s = \hbar k$. Notice that the total energy of this solution as well as the velocity of the solitary wave depends on the value of $q$. Moreover, we have also shown the conservation of momentum (\ref{conservation}). 

It would be interesting to investigate the interaction of Lorentzian solitary waves in order to understand if the density of energy retains the Lorentzian structure after the interaction. Moreover, there are some papers which show the existence of acoustic solitons in what is called nonextensive plasma \cite{mouloud, saini, moufida}, i.e., plasmas with nonextensive distributions of particles. Since the generalised NLSE is connected with nonextensive Boltzmann-Gibbs statistical mechanics, the result of this paper suggests that it may be possible to find Lorentzian acoustic solitons in a nonextensive plasma.

\vspace{1cm} 
\noindent 
{\bf Acknowledgments:}   The author thanks Profs. Fernando D. Nobre, Evaldo M. F. Curado and Constantino Tsallis for useful comments.

\end{document}